# Refractive Index, Its Chromatic Dispersion, and Thermal Coefficients of Four Less Common Glycols

Anastasiya Derkachova[a], Daniel Jakubczyk[a,*], Gennadiy Derkachov[a], Kwasi Nyandey [a,b]

[a] *Institute of Physics, Polish Academy of Sciences, al. Lotników 32/46, 02-668 Warsaw, Poland*

[b] *Laser and Fibre Optics Centre, Department of Physics, School of Physical Sciences, College of Agriculture and Natural Sciences, University of Cape Coast, Cape Coast, Ghana*

** Corresponding author's e-mail: jakub@ifpan.edu.pl*

## Abstract

We report comprehensive measurements of the refractive index as a function of wavelength and temperature for four less commonly studied glycols: pentaethylene glycol, hexaethylene glycol, dipropylene glycol (mixture of isomers), and tripropylene glycol. The measurements cover the spectral range of 0.39–1.07 μm and temperatures from 1 °C to 45 °C. The data were modeled using a two-pole Sellmeier equation, with temperature dependence expressed through wavelength-dependent thermal coefficients. Four fitting models (Sellmeier and Cauchy) with different numbers of parameters were tested; for pentaethylene glycol, results from all models are shown, while for the remaining glycols, only the two-pole Sellmeier fits are presented in tabulated form. Thermal coefficient values for six wavelengths of practical importance are also tabulated. Experimental uncertainties in refractive index, wavelength, and temperature were rigorously evaluated and incorporated into the analysis. The influence of sample purity – including residual water content and manufacturer-reported impurities – was assessed and accounted for in the uncertainty estimates. To our knowledge, this is the first dataset to systematically characterize both chromatic dispersion and thermal variation of the refractive index for these glycols over such a broad spectral and temperature range. The validated fitting equations and parameters are suitable for use in optical modeling, materials characterization, and related applications. All raw data are available in a publicly accessible repository.

## 1. Introduction

Since liquids – particularly in the form of (micro)droplets – can often be readily recognized or characterized by their optical properties, accurate values of their refractive indices at defined wavelengths and temperatures are highly desirable. Also, the field of optofluidics uses liquids of varying refractive indices in microfluidic/optical systems (e.g., tunable fluid waveguides, lenses) – tuning $n(\lambda, T)$ of liquids matters in dynamic optics [1]. However, for less common substances, such data are surprisingly difficult to find in the literature beyond the standard $n_\mathrm{D}^{20}$ value, and the available measurements are often several decades old. Glycols, a subgroup of diols, are no exception, despite their extensive use in numerous industrial applications due to their distinctive physical and chemical properties.

This study focuses on four less commonly studied glycols (for IUPAC names and CAS numbers see Section 2.1): pentaethylene glycol (PEG), hexaethylene glycol (HEG), dipropylene glycol (DPG), and tripropylene glycol (TPG). We provide detailed measurements of their refractive indices, including chromatic dispersion and thermal coefficients, across the visible and near-infrared spectral range (0.39–1.07 μm) and at temperatures from 1°C to 45°C. The work extends our previous research on the optical properties of more common glycols and glycerol [2].

Pentaethylene and hexaethylene glycols, both members of the polyether series, are used in the synthesis of surfactants, lubricants, and polymer intermediates (see e.g. [3]), as well as in heat-transfer and antifreeze formulations. Their physicochemical stability, low volatility, and high boiling points make them suitable for demanding applications such as coatings, adhesives, and high-temperature fluids. Dipropylene and tripropylene glycols are widely employed as solvents, plasticizers, and humectants in coatings, resins, and personal care products. They are also important constituents of hydraulic and de-icing fluids and serve as intermediates in the manufacture of polyurethane and polyester resins (see e.g. [4]). Both these compounds are produced on an industrial scale.

Because glycols are hygroscopic, accurate refractive index measurements require special care to prevent contamination with water. In our work, such precautions were systematically applied, and the potential effects of residual water and synthesis-related impurities on sample purity are discussed in Section 2.3.

We believe that the primary value of our work lies not so much in achieving ultimate pointwise accuracy as in providing a consistent and self-consistent set of refractive-index measurements across a broad ($\lambda$, $T$) domain, enabling reliable fitting and visualization of the $n(\lambda, T)$ surface. These results are compared with the sparse literature data in Section 4 and illustrated in Figs. 5–9, highlighting the completeness and coherence of our dataset relative to previously available values.

## 2. Measurement Methodology

The core of the setup for refractive index measurement consisted of a modified Abbe refractometer (AR-4, Müller). To measure the chromatic dispersion of the refractive index, the compensator (comprising two Amici prisms) was set to maximum dispersion, which consequently required rescaling the refractometer. We used formula (6) from [5], carefully calibrated with water data [6] over the whole accessible ($\lambda$, $T$) range to match our setup.

The light at the desired wavelength was provided by a monochromator (SPM1, Carl Zeiss Jena [7]) illuminated by a halogen lamp (H1 automotive bulb). The monochromator was calibrated *in situ* using a compact grating spectrometer (USB4000, Ocean Optics; 1.4 nm resolution). The data points are not evenly spaced with respect to wavelength because the scale on the prism rotation knob does not translate linearly to wavelength. This, however, proved ad-

vantageous – denser data points in the steeper region of the dispersion curve allow for a more accurate fit (see Fig. 1).

The refractometer eyepiece was equipped with a 14-bit digital camera (GC651MP, Smartek Vision) coupled with an additional camera objective (f = 6 mm, f/1.6, aberration-corrected including IR range). The optical system was arranged so that the shadow resulting from total internal reflection in the prism, the superimposed crosshair, and the scale were simultaneously within the field of view. The shadow-edge brightness profile was obtained using an in-house MATLAB script [8]. The derivative of this profile exhibits a minimum corresponding to the inflection point of the brightness distribution; this point was taken as the shadow-edge position. The crosshair center was determined at the beginning of each measurement series, and the measurement consisted of adjusting the minimum of the profile derivative to coincide with this center.

The AR-4 refractometer enables measurements at different temperatures by circulating a liquid of controlled temperature through the prism jacket. The temperature of the circulating fluid was maintained by a custom-built heat exchanger with a local stabilization loop. Because the built-in thermometer readings were significantly offset from the actual temperature in the prism gap, a separate K-type thermocouple was placed directly next to the prism–liquid contact surface. Temperature was measured with a CHY506R electronic thermometer (CHY Firemate Co.) calibrated against a mercury thermometer with traceability to Polish national standards (Calibration Certificate OUM83/88.80/2006), with an accuracy of ±0.1°C. The data points are spaced by 5°C, except for measurements at 1°C – see section 2.3 (Uncertainty analysis).

As the predominant source of error was found to originate from the refractometer's mechanical accuracy and hysteresis, we sought to minimize it by stabilizing the temperature of the entire refractometer body, independently of the prisms. Dry, filtered $N_2$ gas obtained from liquid nitrogen was circulated through the refractometer chassis and housing. The gas temperature was stabilized by passing it through a gas–liquid heat exchanger in a thermal bath controlled to ±0.2°C. Depending on the prism temperature, the bath temperature was set between −5 and 0°C, keeping it significantly lower than that of the prisms. The $N_2$ temperature was continuously monitored at the refractometer inlet and remained stable within ±1°C during each experimental run.

In addition to temperature stabilization, the $N_2$ flow prevented vapor condensation on the optical surfaces and reduced diffusion of atmospheric moisture into the measured liquid. The relative humidity (RH) inside the enclosure was always below 24%, and long-term monitoring confirmed that the refractive index of a sample placed between the prisms remained constant for several hours. Sufficient time was always allowed for complete stabilization of experimental conditions.

The refractometer calibration procedure is described in section 2.3.

Further details of the experimental setup and procedures are provided in [2]. The corresponding 3D model (.skp, 2017 format) and a walk-around video (.mp4) generated from this model are available in the Mendeley Data repository [9].

## 2.1. Sample Materials

The substances investigated in this study include:

- pentaethylene glycol (IUPAC Name 3,6,9,12-tetraoxatetradecane-1,14-diol, CAS 4792-15-8; 98.2%, Alfa Aesar, lot 10207595 and 97.0%, abcr GmbH, lot 1509179),
- hexaethylene glycol (IUPAC Name 3,6,9,12,15-pentaoxaheptadecane-1,17-diol, CAS 2615-15-8, 96% abcr GmbH, lot 1508060),
- dipropylene glycol (a mixture of isomers, IUPAC Names 4-Oxa-2,6-heptandiol/ 4-Oxa-1,7-heptandiol, CAS 25265-71-8, 99.8%, Alfa Aesar, lot 10201331),
- tripropylene glycol (IUPAC Name 2-[2-(2-Hydroxypropoxy)propoxy]-1-propanol, CAS 24800-44-0, 99.85%, Alfa Aesar, lot H25W018).

All substances were stored under vacuum and the exposure to ambient air was minimized during sample transfer to the refractometer, to prevent contamination or degradation.

## 2.2. Data processing

As we showed in our work for other glycols and glycerol [2], here also the dependence of the refractive index $n$ on the wavelength $\lambda$ and the temperature $T$ could be decomposed – within our measurement accuracy – into two terms:

$$n(\lambda, T) = n(\lambda, 20) + \frac{\mathrm{d}n(\lambda, T)}{\mathrm{d}T}(T - 20)\ . \tag{1}$$

The first term bears only the dependence on $\lambda$ and corresponds to $T = 20$ºC, while the second term is linear in $T$ (here in ºC), within our range of temperatures, and non-linear in $\lambda$. The experimental data was fitted with equation (1) stepwise. The first term was fitted with a two-pole Sellmeier equation – the most common dispersion formula (see e.g. [10] and [11] for physical model):

$$n^2(\lambda, 20) = A + \frac{B_{\mathrm{IR}}\lambda^2}{\lambda^2 - C_{\mathrm{IR}}} + \frac{B_{\mathrm{UV}}\lambda^2}{\lambda^2 - C_{\mathrm{UV}}}\ , \tag{2}$$

where $A$ describes the short-wavelength absorption contributions to $n$ at longer wavelengths, while $B_{\mathrm{IR}}$ and $B_{\mathrm{UV}}$ are absorption resonance strengths at wavelengths $C_{\mathrm{IR}}^{½}$ and $C_{\mathrm{UV}}^{½}$ respectively. The thermal coefficient in the second term could be sensibly fitted with a function of a similar class (compare versus model in [12]):

$$\frac{\mathrm{d}n(\lambda, T)}{\mathrm{d}T} = A_{\mathrm{T}} + \frac{B_{\mathrm{T}}}{\lambda - C_{\mathrm{T}}}\ \ , \tag{3}$$

where $A_{\mathrm{T}}$ is associated mainly with thermal expansivity (density change) of the liquid and $B_{\mathrm{T}}$ and $C_{\mathrm{T}}$ parameters play similar roles as in equation (2).

The coefficients of equations (2) and (3) were obtained by fitting the equations to the measured refractive-index data using the damped least-squares (Levenberg–Marquardt) algorithm in Origin 2021 (OriginLab Corp.). Parameter uncertainties (1σ, 68% confidence) were computed using the Asymptotic-Symmetry-based method, which derives confidence intervals from the covariance matrix under the assumption of local symmetry of the likelihood surface. Experimental-input uncertainties (see section 2.3) were not propagated into these values.

Curves generated using these formulas are shown together with the experimental data in Figs. 1 and 6–9 for the room temperature (20°C) as well as the most extreme temperatures: 1°C, 5°C, and 45°C. In most cases, the discrepancies at 1°C are the largest (see the analysis in section 2.3), but they remain within the estimated limits of error.

Since the second and third terms in equation (2) are mathematically identical, fitting the experimental data with this equation can easily lead to overparameterization, which we indeed observed. To mitigate this, we initialized the parameters $A$, $C_{IR}$ and $C_{UV}$ within the ranges defined by the physical interpretation of the Sellmeier model and iteratively kept one of them fixed in each fitting step to ensure that all parameters remained within those physically meaningful regions.

We have also fitted our experimental data using three other equations. Here, we present the results for PEG, including the fit parameters and their uncertainties:

- 1-pole Sellmeier: $n^2(\lambda, 20) = A + \frac{B\lambda^2}{\lambda^2 - C}$ ; $A$=0.8±0.2, $B$=1.4±0.2, $C$ =0.009±0.001
- 3-term Cauchy: $n(\lambda, 20) = A_1 + \frac{A_2}{\lambda^2} + \frac{A_3}{\lambda^4}$ ; $A_1$=1.45068±7×10$^{-5}$, $A_2$=0.00402±4×10$^{-5}$, $A_3$=3.1×10$^{-5}$±6×10$^{-6}$
- 5-term Cauchy (see e.g. [5]): $n(\lambda, 20) = A_0 + A_1\lambda^2 + \frac{A_2}{\lambda^2} + \frac{A_3}{\lambda^4} + \frac{A_4}{\lambda^6}$ ; $A_0$=1.4519±1×10$^{-4}$, $A_1$=-0.0011±1×10$^{-4}$, $A_2$=0.00378±4×10$^{-5}$, $A_3$=8×10$^{-17}$±0, $A_4$=7.3×10$^{-6}$±5×10$^{-7}$.

The first two models yielded fits with visible but small discrepancies relative to the 2-pole Sellmeier model in the NIR and/or violet regions, as shown in Fig. 1. This behavior is well known (see, e.g., [10] and references therein). The 5-term Cauchy model produced results very similar to those of the 2-pole Sellmeier equation (both containing five parameters); the minute differences are discernible only in the magnified view shown in the right panel of Fig. 1 (again, see [10]). The fitting uncertainties obtained for the Cauchy models are generally smaller than those of the Sellmeier models, and direct comparison between different model types may therefore be misleading.

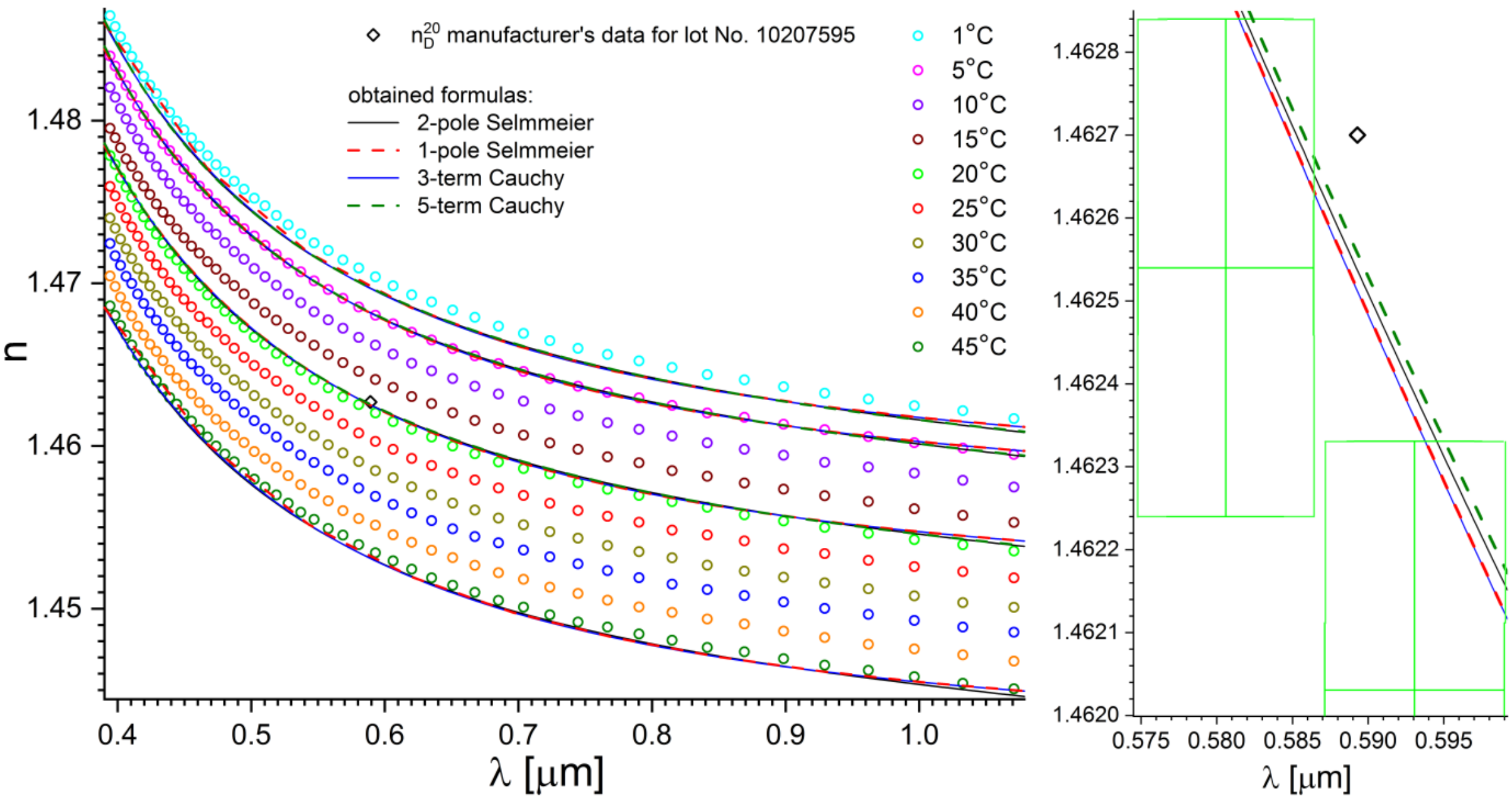

Fig. 1. Comparison of fits obtained using different models for PEG. Error bars – experimental data points uncertainties – are altogether omitted in the left panel for clarity (they are presented in Fig. 6). Small deviations between the models are visible only in the NIR and/or violet regions. The right panel shows a magnified view of the region around $n_D^{20}$, highlighting the subtle differences between the fits. Error bars for $n$ and λ, but not for $T$, are shown there for the two data points closest to $n_D^{20}$.

## 2.3. Uncertainty analysis

Before each measurement series at a given temperature, the refractometer was calibrated with distilled water using recognized reference data [6] for the sodium D-line and for the most extreme wavelengths accessible in our experiments (390 and 1070 nm). While calibration with liquid water is naturally impossible below 0°C, it becomes unreliable even below 5°C, since d$n$/d$T$ for water approaches zero as $T \rightarrow 0$°C. Therefore, for measurements at 1°C, calibration was performed using our previously obtained data [2] for ethylene glycol (MEG). This, however, reduced the refractive-index accuracy at 1°C by approximately a factor of two. In our earlier measurements [2], this deterioration was not apparent and thus went unnoticed; in the present series, however, it became clearly visible, prompting a re-evaluation of the measurement accuracy, now reflected by distinct error bars.

In this context, it appears natural to consider the use of another nonaqueous certified standard for improved accuracy. However, to the best of our knowledge, no such reference liquids have refractive-index values experimentally verified at low temperatures. Instead, it is generally assumed that measurements performed at higher temperatures can be corrected using temperature coefficients. These, however, are typically derived over a different temperature range. Moreover, the temperature coefficient is by default assumed to be linear and wavelength-independent, which introduces unquantified systematic errors. For these reasons, we consider it preferable to use MEG at 1°C. We have carefully measured it in previous experiments, and a substantial body of literature data exists for it – although again not below 15°C. Measuring refractive index at 0°C and below remains an even more challenging task.

A similar issue arises for calibration above 45°C. In a dry nitrogen environment at elevated temperatures, water evaporates from between the prisms during calibration, rendering the procedure unreliable. Consequently, calibration with another standard liquid of lower volatility would be required. It seems that this could, in principle, be achieved with better results than for the low-temperature case discussed above. In this way, the temperature range of industrial relevance (e.g., heat transfer fluids) could be covered, and we intend to pursue this direction in future studies. However, although the AR-4 refractometer is nominally capable of operating up to 70°C, we found ~60°C to be the maximum safe operating temperature of the setup, due to the softening and deformation of plastic components.

After each measurement series at a given temperature, the calibration of the refractometer was verified with the same calibration liquid. The measurements, for which the verification failed, were simply discarded.

The absolute precision of refractive index measurements with an AR-4 Abbe refractometer can be estimated as $\pm 1 \times 10^{-4}$. However, the actual accuracy of the instrument is lower due to mechanical hysteresis and limited long-term thermal stability. We estimate the overall accuracy of the apparatus to be $\Delta n_{apar} = \pm 3\times10^{-4}$. The total uncertainty of the refractive-index measurements may also need to include additional contributions arising from the limited purity of the investigated materials.

The residual refractive-index uncertainty, $\Delta n_{puri}$, arising from the limited purity of the liquid (water and/or other impurities, as stated – if available – by the manufacturer for each lot used) was estimated using the linear volume-fraction model as follows:

- DPG: $\Delta n_{puri} = +4\times10^{-5}$ – influence of water only (0.04%, Karl-Fisher). The remaining 0.16% of impurities (determined by GC with silyl derivatization) were not identified or named by the manufacturer in the Certificate of Analysis for the lot. The "+" sign indicates that the refractive index of the pure substance is expected to be slightly higher than the measured value;
- TPG: $\Delta n_{puri} = \pm 2\times10^{-5}$ – influence of water (0.016%) and tetrapropylene glycol (0.13%, as stated in the Certificate of Analysis for the lot; $n_{tetraPG} = 1.455$ [13]), hence "±" sign applies here;
- PEG: $\Delta n_{puri} = +3\times10^{-4}$ – two lots of 98.2% and 97% purity were used, yielding no perceptible difference in results. The impurity composition was not specified in the Certificate of Analysis. A 0.2 vol% water content was assumed as impurity; in [14], PEG of comparable purity (as stated by the manufacturer) was vacuum-dried to 0.12 wt%, consistent with our procedure of storing the liquids under vacuum. Possible contamination with HEG or tetraethylene glycol should not introduce any significant $\Delta n_{puri}$, as both have refractive indices close to that of PEG;
- HEG: $\Delta n_{puri} = +3\times10^{-4}$ – based on data from [14] and assuming again 0.2 vol% of water. No Certificate of Analysis was provided by the manufacturer.

It can be seen that, for DPG and TPG, the residual content of water and other impurities contributes very little to the overall refractive-index measurement uncertainty. In contrast, for PEG and HEG, the purity-related uncertainties are comparable to the accuracy of the apparatus, but they do not affect the results symmetrically – possible water contamination may cause the measured refractive index to be slightly underestimated.

The total uncertainty of the refractive-index measurements $\Delta n = \Delta n_{apar} + \Delta n_{puri}$ is represented by vertical error bars in Figs. 6–9. For PEG and HEG (Figs. 8–9), these error bars are clearly asymmetric.

It must be also kept in mind that a significant error can be introduced into the refractive index measurement, if a hygroscopic sample is exposed to a humid ambient atmosphere for a longer period. The equilibrium water content in a hygroscopic liquid, such like a glycol, can increase to several percent (compare Fig. 18 in [15]), which would lead to a decrease of the refractive index of the order of $10^{-2}$ (as obtained e.g. with the linear volume fraction model). This would count as a very significant error, which renders the result of little value. Thus, extreme care was taken to avoid any prolonged contact of samples with the ambient atmosphere or water-saturated plastic of syringes used for the liquid transfer. Also, in view of that, the liquids and the syringes were vacuum-dried at room temperature and then stored under vacuum. This procedure removed also other compounds of higher volatility. The possible water intake during the measurement run was carefully verified in a separate experiment and found definitely negligible.

The maximum uncertainty in wavelength determination was dictated by the calibration accuracy of the monochromator–halogen lamp system. The precision of setting the wavelength is better than ±0.75 nm. However, in order to have sufficiently bright illumination we opened the slits wide, which yielded spectrally non-uniform illumination. Spectral profiles differ versus the central wavelength used – the shorter wavelength the narrower the profile: from 14 nm HWHM at 1.071 μm to 3 nm HWHM at 0.394 μm, as we measured in our setup [2] – see Fig. 2. This led to the total wavelength uncertainty of ~1%, which is presented as horizontal error bars in Figs. 6–9. Due to the character of the dispersion curve, the influence of wavelength uncertainty on the fitted Sellmeier curve is comparable to that introduced by refractive index uncertainty.

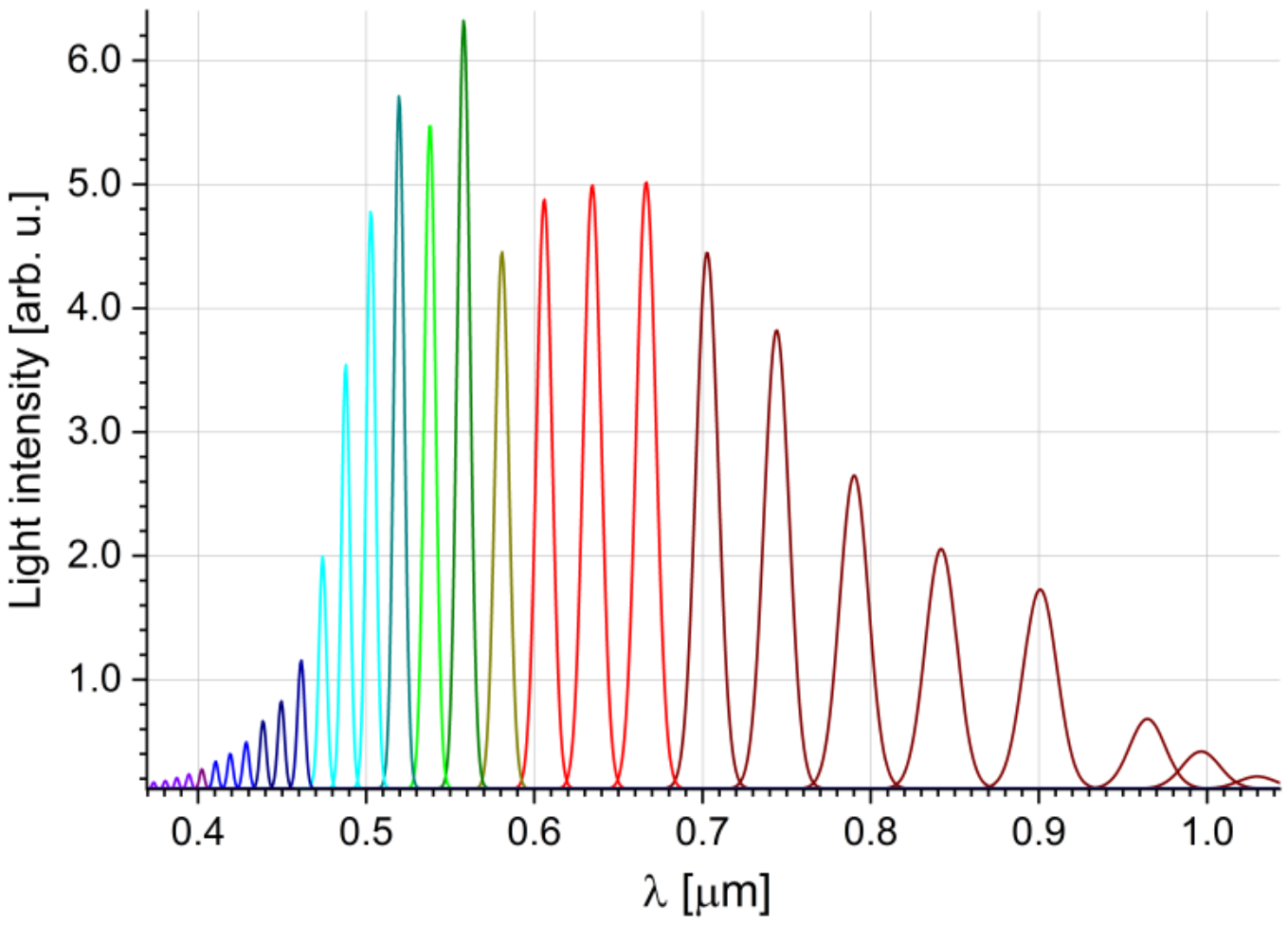

Fig. 2. Light spectra at 29 of the 50 wavelengths sequentially selected by the experimenter using the monochromator in each experimental series. The illumination was provided by a 100 W / 12 V halogen bulb and directed to the refractometer.

The temperature of the investigated liquid was continuously monitored during the measurement runs with a K-type thermocouple placed directly next to the gap between the prisms, containing the liquid under the study. Such arrangement was necessary since we observed a temperature difference between prisms-liquid contact surface and the circulating liquid up to 2°C (depending on the set temperature of the refractometer). The uncertainty of these measurements consists mainly of two parts: (i) accuracy of the local temperature measurement with a thermocouple, and (ii) the temperature gradients that may arise in the liquid between the prisms. We measured the temperature difference between the gap's edge (K-type thermocouple) and the center of the gap (0.17-mm-thin T-type thermocouple, TT-T-40-SLE by Omega) – see Fig. 3 – for various experimental conditions. It was never greater than ±0.3°C.The accuracy of (both) thermocouples used was traced to the officially calibrated thermometer and estimated as ±0.2°C. Thus, the overall accuracy of temperature measurements can be estimated as ±0.5°C. This uncertainty is indicated by the thickness of the grey lines in Figs. 6–9.

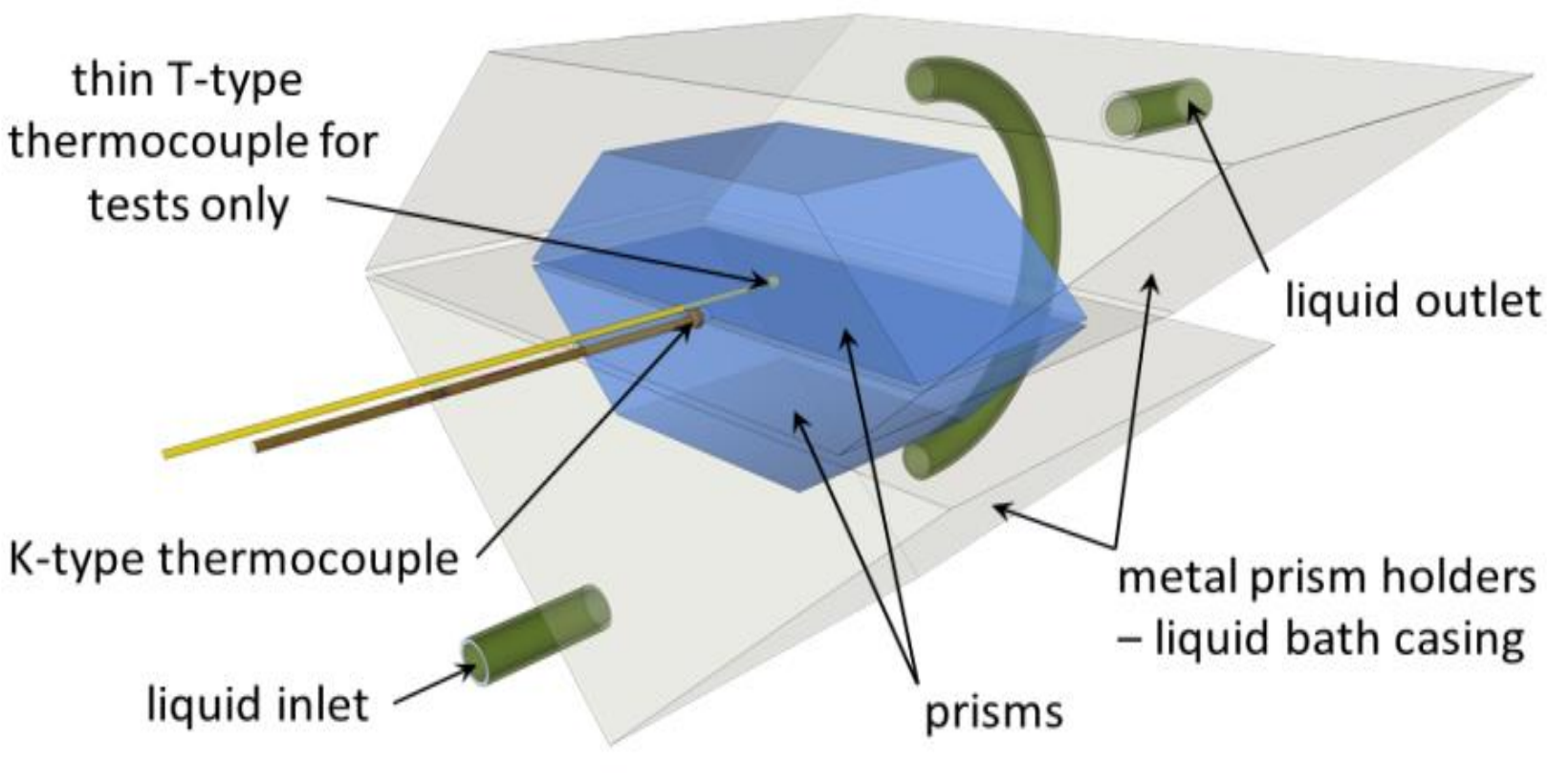

Fig. 3. Details of refractometer prisms thermal bath and thermocouple placement. The K-type thermocouple performs constant temperature monitoring during measurements, while the thin T-type thermocouple was only used to check the temperature gradient across the gap between the prisms.

Since d$n$/d$T$ was found with a linear fit from the experimental data for each experimental $\lambda$ value, the uncertainty of d$n$/d$T$ was also provided by the fitting procedure as the standard error. The extent of these uncertainties is visualized in Fig. 4 for DPG.

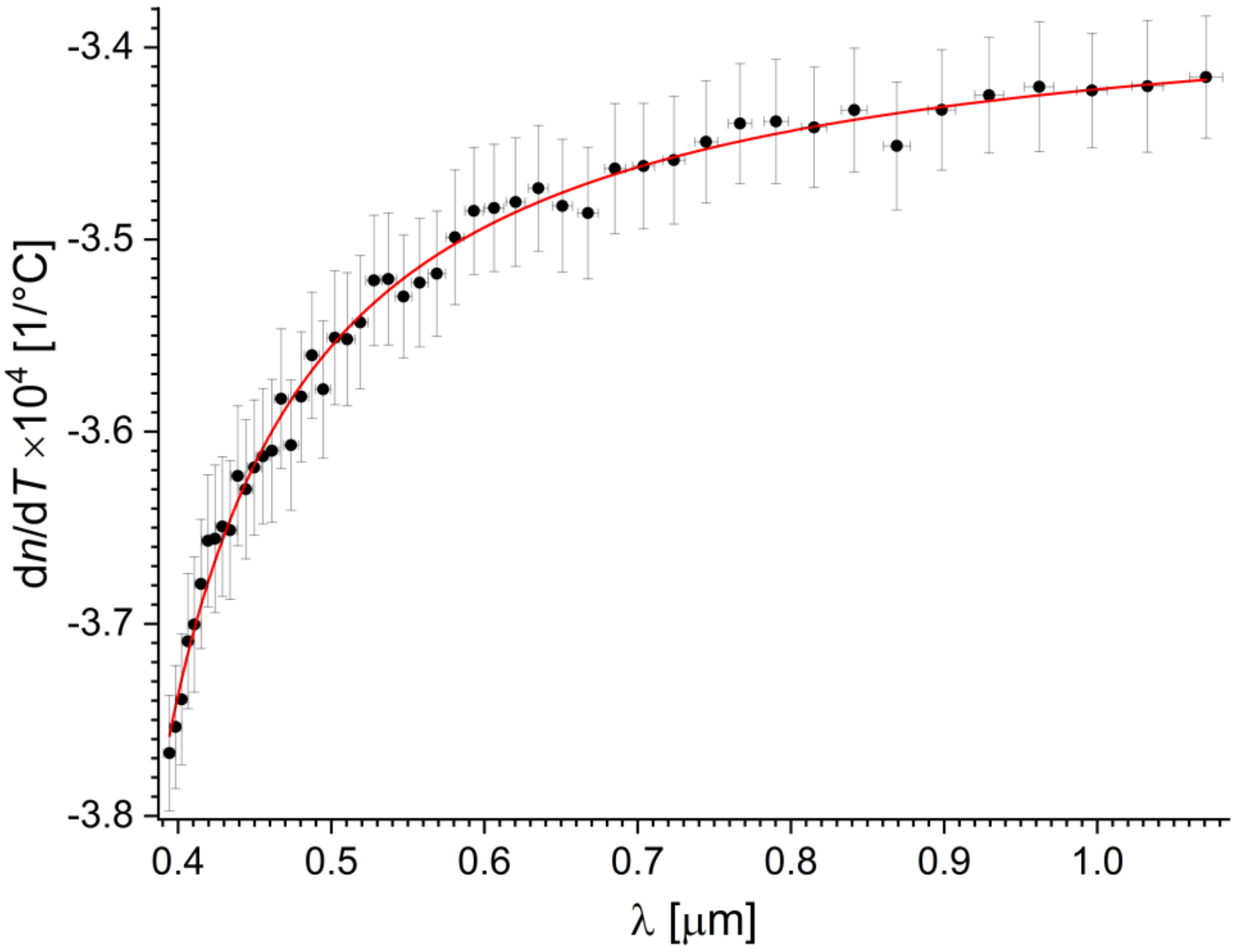

Fig. 4. Thermal coefficients d$n$/d$T$ for dipropylene glycol (DPG), obtained by linear fits to the refractive index data at each measured wavelength. Vertical error bars represent the standard errors from the fitting procedure. Horizontal error bars indicate the estimated uncertainties in wavelength.

## 3. Results – Chromatic Dispersion and Thermal Coefficients of Refractive Index

The raw data are available in the Mendeley Data repository [9] in csv format. The measured refractive indices of PEG, HEG, DPG, and TPG show a monotonic decrease with increasing wavelength and temperature (which is consistent with the behavior we observed in other glycols and glycerol [2]) – see 3D Fig. 5 for DPG as an example. Chromatic dispersion, quantified as the rate of change of refractive index with wavelength, is most pronounced in the violet region and diminishes towards the near-infrared region. Figures depicting the refractive index as a function of wavelength and temperature for each substance are included in this paper (Figs. 6–9). The fitted parameters of equations (2) and (3) for the four liquids are collected in Tables 1 and 2. The thermal coefficients (d$n$/d$T$) were negative for all substances, and exhibited monotonic, non-linear dependence of the wavelength – see Fig. 4 for illustration (compare [12]). The coefficients for 6 wavelengths of practical importance are presented in Table 3 for all 4 glycols together with their uncertainties. The uncertainties were propagated using the exact differential method applied to equation (3).

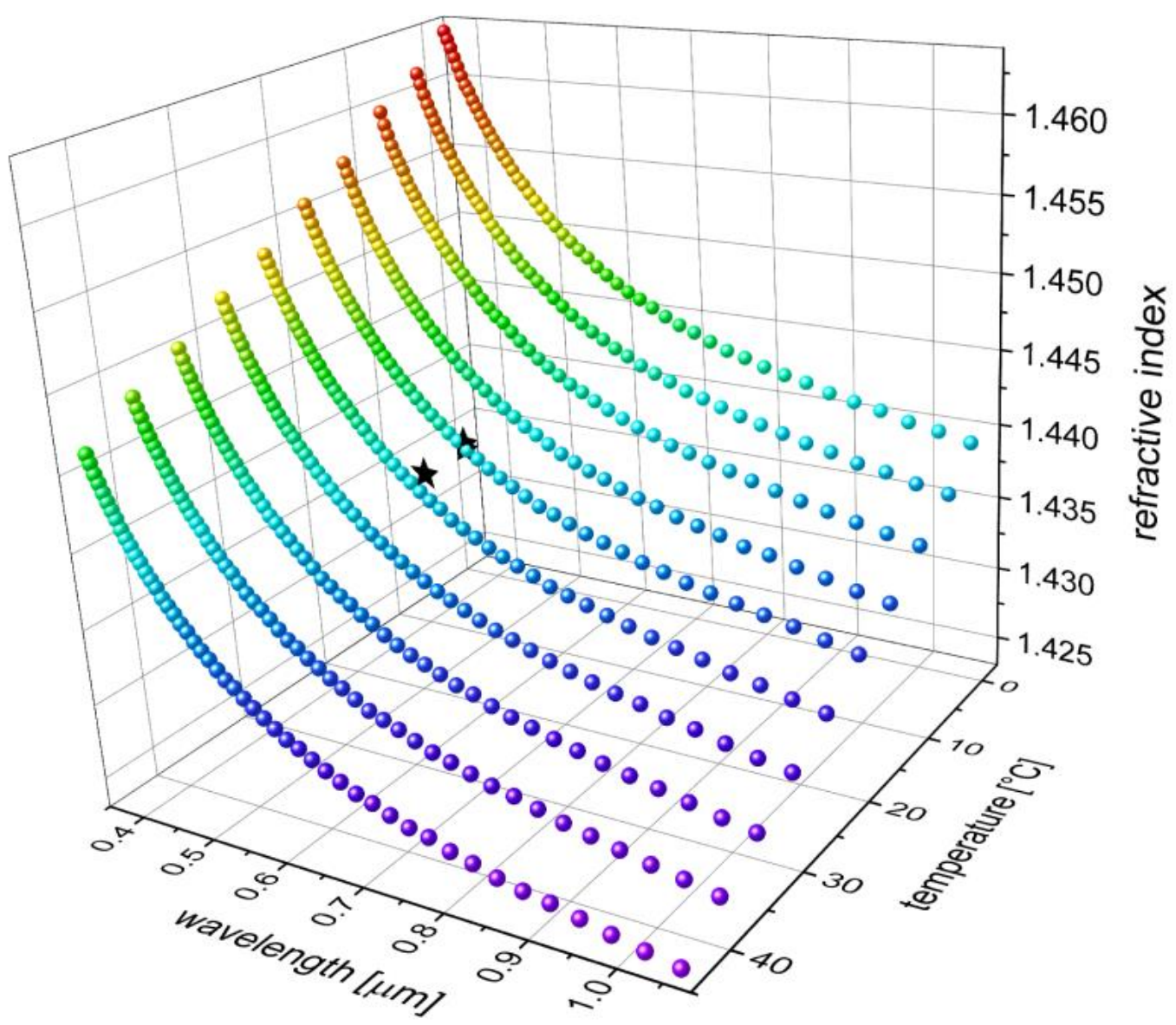

Fig. 5. Refractive index of dipropylene glycol (mixture of isomers) versus wavelength and temperature. The value of measured refractive index is additionally color-coded for clarity. The available literature data is shown as black stars. For the details on literature data, error bars and Sellmeier formula fits, see Fig. 6.

Table 1. Sellmeier equation coefficients for $\lambda$ in μm, found from the presented experiments. Uncertainties represent standard errors reported by the fitting procedure – see text.

| glycols: | Sellmeier equation coefficients | | | | |
|---|---|---|---|---|---|
| | $A$ | $B_{IR}$ | $C_{IR}$ | $B_{UV}$ | $C_{UV}$ |
| pentaethylene | 1.60 ± 0.04 | 0.0096 ± 9×10$^{-4}$ | 4.13± 0.05 | 0.51 ± 0.04 | 0.0202 ± 0.0012 |
| hexaethylene | 1.649 ± 0.035 | 0.0053 ± 6×10$^{-4}$ | 3.01 ± 0.05 | 0.463 ± 0.035 | 0.0222 ± 0.0014 |
| dipropylene | 1.64 ± 0.03 | 0.021 ± 0.002 | 6.07±0.07 | 0.40 ± 0.03 | 0.0232 ± 0.0015 |
| tripropylene | 1.60 ± 0.03 | 0.0041 ± 4×10$^{-4}$ | 2.61 ± 0.04 | 0.46 ± 0.03 | 0.0213 ± 0.0013 |

Table 2. Thermal coefficients for $\lambda$ in μm found from the presented experiments. Uncertainties represent standard errors reported by the fitting procedure – see text.

| glycols: | thermal coefficients | | |
|---|---|---|---|
| | $A_T$ | $B_T$ | $C_T$ |
| pentaethylene | -3.631×10$^{-4}$ ± 6×10$^{-7}$ | -4.78×10$^{-6}$ ± 6×10$^{-8}$ | 0.25 ± 0.03 |
| hexaethylene | -3.486×10$^{-4}$ ± 3×10$^{-7}$ | -5.24×10$^{-6}$ ± 3×10$^{-8}$ | 0.243 ± 0.015 |
| dipropylene | -3.367×10$^{-4}$ ± 3×10$^{-7}$ | -3.84×10$^{-6}$ ± 5×10$^{-8}$ | 0.296 ± 0.006 |
| tripropylene | - 3.512×10$^{-4}$ ± 3×10$^{-7}$ | -4.54×10$^{-6}$ ± 3×10$^{-8}$ | 0.267 ± 0.016 |

Table 3. Thermal coefficients d$n$/d$T$ [1/°C] calculated with equation (3) for six wavelengths of practical importance. Uncertainties were propagated using the exact differential method applied to equation (3).

| glycols: | $\lambda$ [μm] | | | | | |
|---|---|---|---|---|---|---|
| | 0.405 | 0.532 | 0.589 | 0.633 | 0.808 | 1.064 |
| pentaethylene | - 3.94×10$^{-4}$ ± 7×10$^{-6}$ | -3.80×10$^{-4}$ ± 3×10$^{-6}$ | -3.77×10$^{-4}$ ± 2×10$^{-6}$ | -3.76×10$^{-4}$ ± 2×10$^{-6}$ | -3.72×10$^{-4}$ ± 1×10$^{-6}$ | -3.690×10$^{-4}$ ± 9×10$^{-7}$ |
| hexaethylene | -3.81×10$^{-4}$ ± 4×10$^{-6}$ | -3.67×10$^{-4}$ ± 2×10$^{-6}$ | -3.67×10$^{-4}$ ± 1×10$^{-6}$ | -3.620×10$^{-4}$ ± 9×10$^{-7}$ | -3.579×10$^{-4}$ ± 6×10$^{-7}$ | -3.550×10$^{-4}$ ± 5×10$^{-7}$ |
| dipropylene | -3.72×10$^{-4}$ ± 3×10$^{-6}$ | -3.530×10$^{-4}$ ± 9×10$^{-7}$ | -3.498×10$^{-4}$ ± 7×10$^{-7}$ | -3.481×10$^{-4}$ ± 7×10$^{-7}$ | -3.442×10$^{-4}$ ± 5×10$^{-7}$ | -3.417×10$^{-4}$ ± 4×10$^{-7}$ |
| tripropylene | - 3.84×10$^{-4}$ ± 4×10$^{-6}$ | -3.68×10$^{-4}$ ± 1×10$^{-6}$ | -3.65×10$^{-4}$ ± 1×10$^{-6}$ | -3.636×10$^{-4}$ ± 9×10$^{-7}$ | -3.596×10$^{-6}$ ± 6×10$^{-7}$ | -3.569×10$^{-6}$ ± 5×10$^{-7}$ |

## 4. Discussion and Comparison with Literature Data

For each of the studied glycols, for a given λ and *T*, *n*(λ, *T*) increases monotonically with the molar mass (similarly, but much less pronouncedly with density). We see no such obvious relation for (d*n*/d*T*)(λ). The negative value of d*n*/d*T* is considered to be primarily associated with normal thermal expansion of the liquid. The lowest absolute value of d*n*/d*T* was consistently exhibited by dipropylene glycol and the highest – by pentaethylene glycol.

This work provides a comprehensive dataset, including temperature-dependent chromatic dispersion, which has not been reported for any of the substances investigated here. The refractive indices measured in this study are generally in good agreement with sparse previously published values and in general comparable to those reported for other glycols.

For PEG and HEG there are several $n_D^{20}$ entries in the Landolt-Börnstein Numerical Data and Functional Relationships in Science and Technology compendium [16]. Most of them fall within our estimated error limits – see Figs. 8–9, and there are few significant outliers. For DPG and TPG there are no entries in [16], apart from a single $n_D^{60}$ entry, which is beyond our measurement range. For DPG and PEG, the manufacturer (Alfa Aesar – Thermo Fisher Scientific) provided $n_D^{20}$ data for the specific lots used, which are in very good agreement with our results.

Measurements at other temperatures for the sodium D-line are considerably rarer for any of the studied liquids. Only a handful of such data points can be found in such compendia like

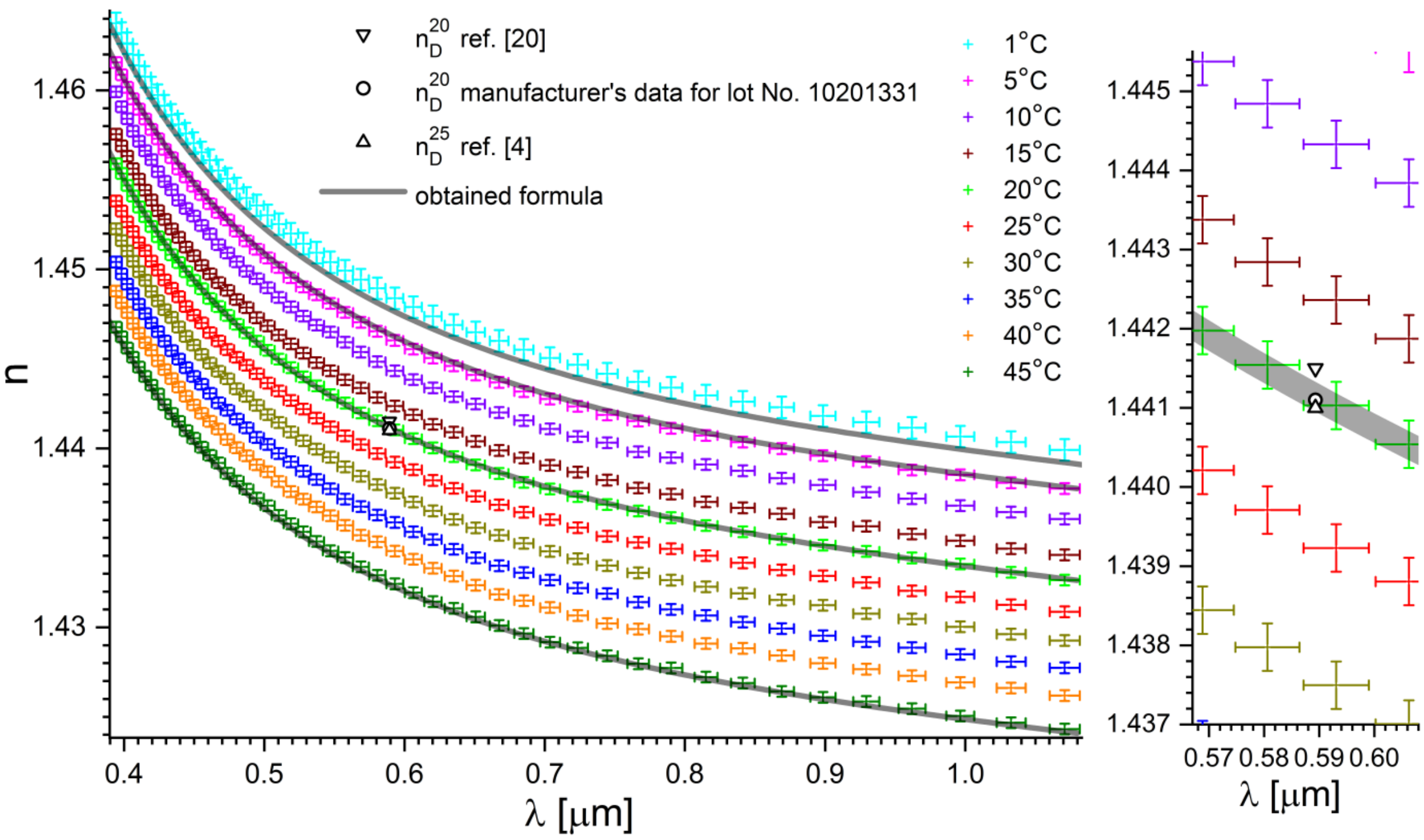

Fig. 6. Left panel: refractive index of dipropylene glycol (DPG) versus wavelength for a range of temperatures (color-coded). Grey solid lines represent selected fits obtained using equations (1)–(3) with parameters from Tables 1 and 2. Uncertainties in *n* and λ are shown as vertical and horizontal error bars, respectively, while those in *T* are indicated by the thickness of the grey lines. The manufacturers' datum for the lot is presented as a black hollow circle, while the literature data known to us are shown as black hollow up-triangle [4] and down-triangle [20]. The region of these data is shown magnified in the right panel.

Landolt-Börnstein – citing [17–19], Ullmann's Encyclopedia of Industrial Chemistry [20] or Kirk-Othmer Encyclopedia of Chemical Technology [4]. These data usually fall outside our estimated error limits (and usually below) – see Figs. 6–9, except for that from [20] – hollow down-triangle in Fig. 7 and [19] – hollow up-triangle in Fig. 9.

We found no data in the literature on chromatic dispersion for any of the considered substances.

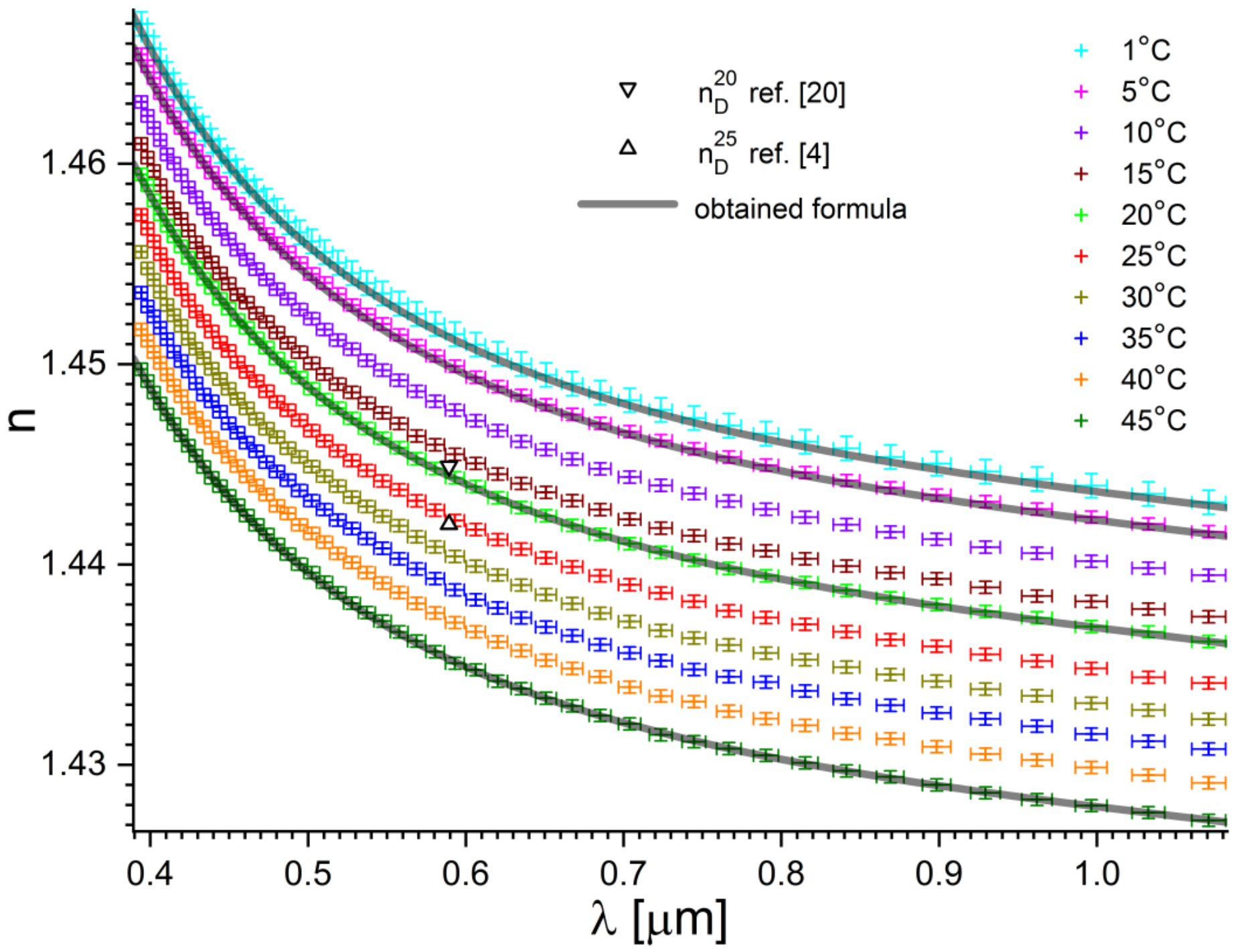

Fig. 7. Refractive index of tripropylene glycol (TPG) versus wavelength for a range of temperatures (color-coded). Grey solid lines represent selected fits obtained using equations (1)–(3) with parameters from Tables 1 and 2. Uncertainties in *n* and λ are shown as vertical and horizontal error bars, respectively, while those in *T* are indicated by the thickness of the grey lines. The manufacturer did not provide the refractive index value for the lot. The literature data known to us are shown as black hollow up-triangle [4] and down-triangle [20].

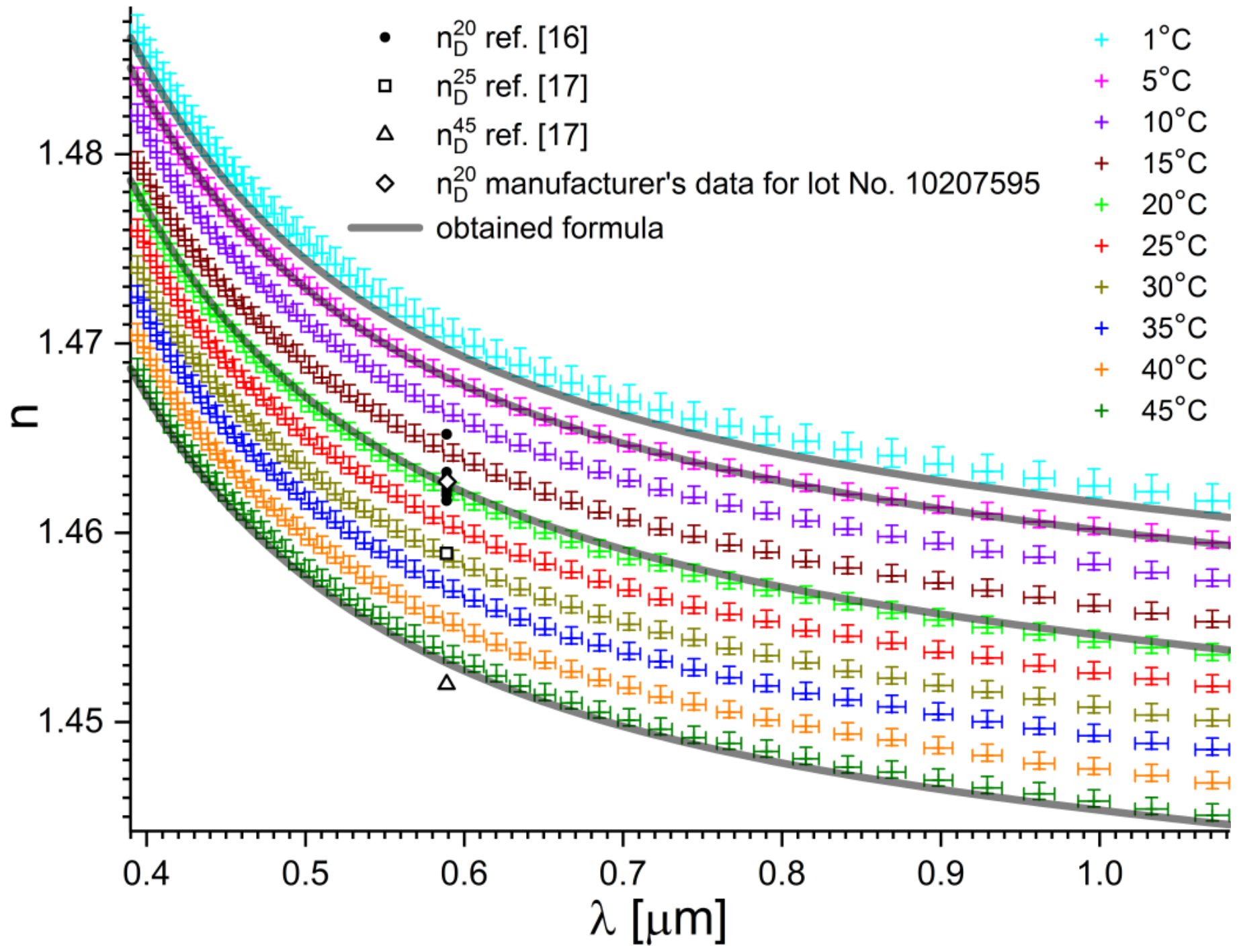

Fig. 8. Refractive index of pentaethylene glycol (PEG) versus wavelength for a range of temperatures (color-coded). Grey solid lines represent selected fits obtained using equations (1)–(3) with parameters from Tables 1 and 2. Uncertainties in $n$ and λ are shown as vertical (asymmetrical) and horizontal error bars, respectively, while those in $T$ are indicated by the thickness of the grey lines. The manufacturers' datum for the lot is presented as a black hollow diamond. The literature data known to us are shown as black dots – compiled $n_D^{20}$ data from [16], black hollow square and up-triangle [17].

## 5. Conclusions

We have measured and analyzed the refractive index of four less commonly studied glycols – pentaethylene glycol, hexaethylene glycol, dipropylene glycol (mixture of isomers), and tripropylene glycol – over a spectral range of 0.39–1.07 μm and temperatures from 1°C to 45°C. The chromatic and thermal behavior of the refractive index was modeled using a two-pole Sellmeier equation and wavelength-dependent thermal coefficients. Fit parameters are provided in tabulated form, along with comprehensive uncertainty estimates derived from experimental precision, temperature control, wavelength setting, and sample purity considerations.

This dataset systematically extends the refractive index reference data available for hygroscopic organic liquids. To our knowledge, it is the first to provide validated, temperature- and wavelength-resolved dispersion data for the studied glycols. The results are suitable for use in optical modeling, calibration, and the interpretation of experimental data involving these liquids. All measurements and fitted values are accompanied by uncertainty analysis, and the full dataset is available in a publicly accessible repository [9].

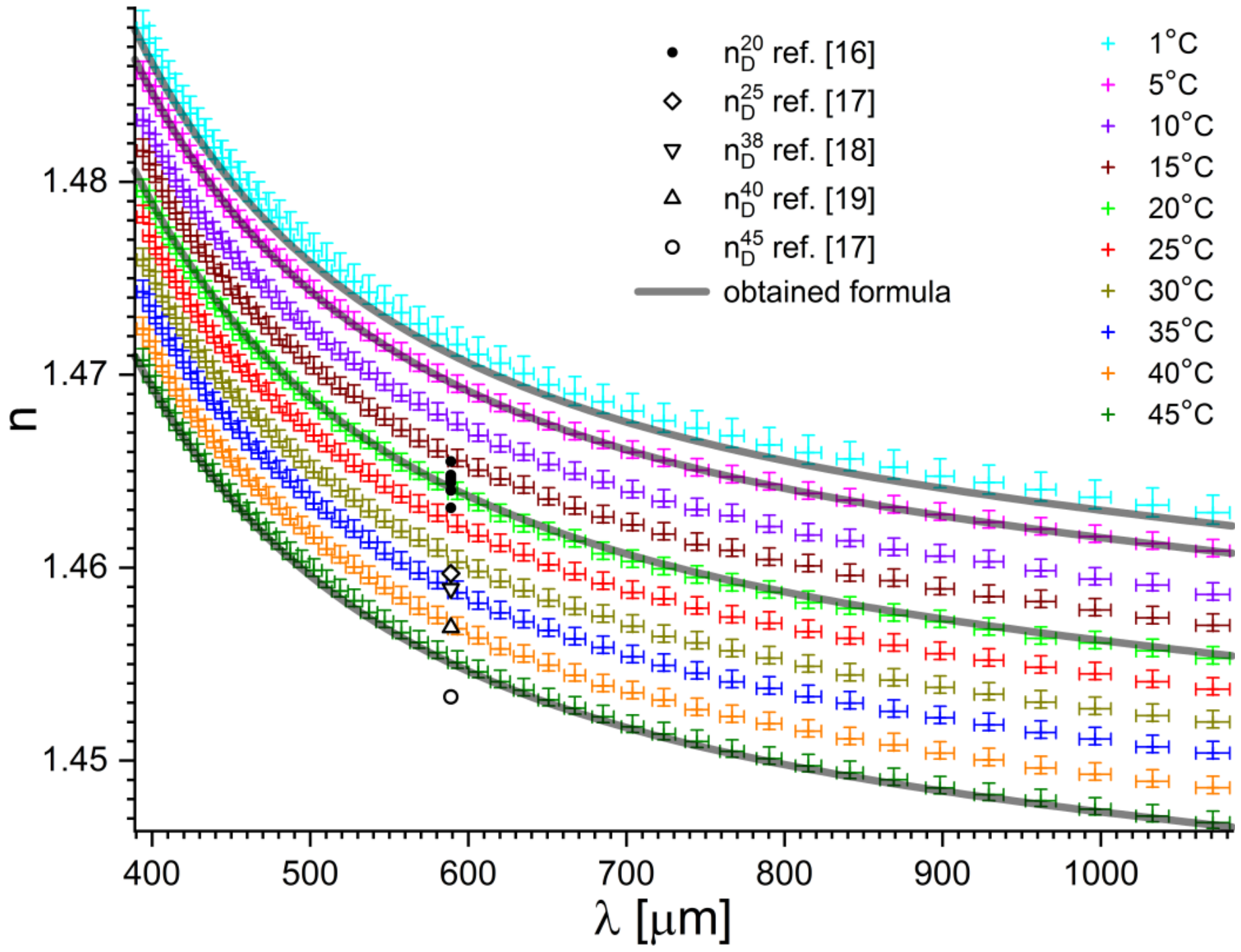

Fig. 9. Refractive index of hexaethylene glycol (HEG) versus wavelength for a range of temperatures (color-coded). Grey solid lines represent selected fits obtained using equations (1)–(3) with parameters from Tables 1 and 2. Uncertainties in $n$ and λ are shown as vertical (asymmetrical) and horizontal error bars, respectively, while those in $T$ are indicated by the thickness of the grey lines. The literature data known to us are shown as black dots – compiled $n_D^{20}$ data from [16], black hollow diamond and circle [17], down-triangle [18] and up-triangle [19].

## 6. Acknowledgment

This research was funded in whole or in part by National Science Centre, Poland, grant 2021/41/B/ST3/00069. For the purpose of Open Access, the author has applied a CC-BY public copyright license to any Author Accepted Manuscript (AAM) version arising from this submission.

## 7. Author Declarations

### 7.1. Conflict of interest

The authors declare that they have no known competing financial interests or personal relationships that could have appeared to influence the work reported in this paper.

## 7.2. Declaration of AI-assisted technologies in the writing process

During the preparation of this work the authors used ChatGPT 5 in order to interactively compose Abstract and Conclusion, and to improve the style of other sections. After using this tool/service, the authors reviewed and edited the content as needed and take full responsibility for the content of the published article.

# 8. Data Availability

The data that support the findings of this study are available in Mendeley Data at https://doi.org/10.17632/gnrpbkhsj3.3

# 9. Referneces